\documentclass[useAMS,usenatbib,usegraphicx]{mn2e}
\title[sausage soliton in the solar atmosphere]{Propagation of sausage soliton in the solar lower atmosphere observed by HINODE/SOT}
\author[T. V. Zaqarashvili et al.]{T. V. Zaqarashvili$^{1,2}$\thanks{E-mail:
teimuraz.zaqarashvili@oeaw.ac.at}, V. Kukhianidze$^{2}$ and M. L.
Khodachenko$^{1}$\\
  $^1$Space Research Institute, Austrian Academy of Sciences, Schmiedlstrasse 6, 8042 Graz, Austria\\
  $^2$Abastumani Astrophysical Observatory at Ilia State University, Al Kazbegi ave. 2a, 0160 Tbilisi, Georgia}

\begin{document}

\date{Accepted . Received }

\pagerange{\pageref{firstpage}--\pageref{lastpage}} \pubyear{2010}

\maketitle

\label{firstpage}

\begin{abstract}

Acoustic waves and pulses propagating from the solar photosphere
upwards may quickly develop into shocks due to the rapid decrease of
atmospheric density. However, if they propagate along a magnetic
flux tube, then the nonlinear steepening may be balanced by tube
dispersion effects. This may result in the formation of sausage
soliton. The aim of this letter is to report an observational
evidence of sausage soliton in the solar chromosphere. Time series
of Ca II H line obtained at the solar limb with the Solar Optical
Telescope (SOT) on the board of Hinode is analysed. Observations
show an intensity blob, which propagates from 500 km to 1700 km
above the solar surface with the mean apparent speed of 35 km
s$^{-1}$. The speed is much higher than expected local sound speed,
therefore the blob can not be a simple pressure pulse. The blob
speed, length to width ratio and relative intensity correspond to
slow sausage soliton propagating along a magnetic tube. The blob
width is increased with height corresponding to the magnetic tube
expansion in the stratified atmosphere. Propagation of the intensity
blob can be the first observational evidence of slow sausage soliton
in the solar atmosphere.

\end{abstract}


\section{Introduction}

Energy transport from the solar photosphere towards the corona,
which eventually may lead to coronal heating, is still an open
problem. There are several possible ways of the energy transport:
waves, pulses or electric currents. The energy transport by the
waves has been recently observed through the oscillatory motions of
plasma in the chromosphere  \citep{kukhianidze,zaqarashvili,De Pontieu,Jess,zaqarashvili1}. On the other hand, the dynamic
photosphere may excite pulses due to convective shootings and/or
magnetic reconnections, which then may propagate upwards. Several
kinds of impulsive events are frequently observed on the solar disc:
chromospheric bright grains  \citep{lites},  blinkers
\citep{Harrison} and explosive events \citep{Porter}. Recent
observations by Hinode spacecraft revealed various types of
energetic events such as chromospheric jet-like structures
\citep{Katsukawa,Shibata,Nishizuka} and type II spicules \citep{De Pontieu1}. However, direct
observational evidence of pulse propagation at the solar limb from
the photosphere upwards, to our knowledge, was not reported yet.

Upward propagating pressure pulses may quickly steepen into shocks
due to the rapid decrease of density. However, if the pulses
propagate along magnetic flux tubes, then tube dispersive effects
may prevent the nonlinear steepening. This may lead to the
formation of a soliton, which is a stable structure propagating
without significant change of shape. The formation of sausage
solitons in magnetic tubes first has been suggested by \citet{Roberts1}. Since that, numerous papers addressed the  soliton
formation problem \citep{Roberts,Roberts2,Merzljakov,Merzljakov2,Sahyouni,Ofman,zhugzhda,Nakariakov,ruderman,ballai,erdelyi,ryutova}. Most of the studies consider a sausage
soliton ($m=0$ mode in magnetic tubes), but no observational support
to the theory was reported yet. On the other hand, some observations
suggest the propagation of nonlinear soliton-like kink waves ($m=1$
mode in tubes) identified with moving magnetic features around
sunspots \citep{ryutova}.

Here we report the upward propagation of a pressure blob in time
series of Ca II H line obtained by Hinode/SOT \citep{Tsuneta}.
Estimated parameters of the blob fit with a solution of slow sausage
soliton propagating along a magnetic tube. Therefore, we suggest
that this is the first observational evidence of sausage soliton
propagation in the lower solar atmosphere.

\begin{figure}
\begin{center}
\includegraphics[width=9.3cm]{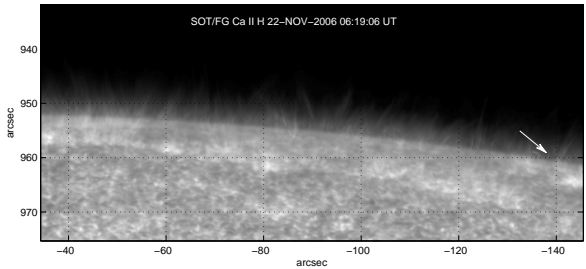}
\end{center}
\caption{Corrected Ca II H image of quiet Sun obtained by
Hinode/SOT. The image was rotated by 90$^0$, therefore the $x$-axis
corresponds to the Solar-$Y$ and the $y$-axis corresponds to the
Solar-$X$. The white arrow shows the place of intensity blob
propagation.
              }
\end{figure}

\section{Observations\label{sec:obs}}

We use Ca II H time series of quiet Sun regions observed by
Hinode/SOT. The spatial resolution of observation reaches 0.2 arc
sec (150 km) and the pixel size is 0.054 arc sec ($\sim$ 40 km). The
observational sequence run on 22th November, 2006 from 05:57:31 U.T.
to 06:34:57 U.T. The position of the X-centre and Y-centre of slot
are, respectively, 960 arc sec and -90 arc sec, while the X-FOV and
Y-FOV are respectively 56 arc sec  and 112 arc sec. The exposure
time for each image is 0.512 s. The integration time for each step
of time series is uniform and equal to 4.8 s.

We start with the raw (zero level) data, then use the standard SOT
subroutines for calibration. The subroutines can be found in the
SSWIDL software tree\\
(http://sohowww.nascom.nasa.gov/solarsoft/hinode/sot/idl). These
subroutines correct the CCD readout anomalies, bad pixels and
flatfield; subtract the dark pedestal and current and apply the
radiation despiking.

\section{Results}

Analysis of the time sequence between 06:19:01 and 06:19:36 U.T.
clearly shows upward propagating pattern in the form of intensity
blob. Fig. 1 displays the corrected Ca II H image taken at 06:19:06
U.T. (the arrow shows the place of the blob propagation). Fig. 2
shows 8 consecutive images of the sequence (left to right and top to
bottom). The time interval between consecutive images is $\sim$ 5 s.
The blob is located at $\sim$ 500-600 km above the surface (see the
upper left panel of Fig. 2) at the moment of 06:19:01 U.T., then it
gradually propagates upwards. The blob is displaced at $\sim$ 1200
km distance during 35 s, therefore mean apparent propagation speed
is $\sim$ 35 km s$^{-1}$. In the image, the blob propagates with $\sim 35^0$ angle
about the vertical. The propagation angle seems larger on Fig. 2,
but this is due to the limb inclination (see Fig. 1). The blob may
propagate also with some angle about the projected plane, then the
real propagation speed can be higher. For an estimate, we may
suppose the same angle of propagation, i.e. $\sim 35^0$, which gives
the real propagation speed as $\sim 42$ km s$^{-1}$. The ratio
between the blob and background intensities is $\sim$ 1.4.
Therefore, the relative amplitude of the density enhancement is
$\sim$ 0.2. The amplitude of the blob is strong enough and indicates
to its nonlinear character. The strong amplitude of pulse density
excludes the possibility of kink or Alfv\'enic pulse. Therefore, it
should be a pressure pulse, which in magnetic tubes transforms into
a sausage pulse. The ratio of blob length to width can be roughly
estimated. Fig. 3 (upper panel) shows the ratio as a function of
time. In the first 4 images (between 0 and 15 s, which corresponds
to the location of the blob at lower heights) the ratio is
approximately 3.5 and later it gradually reduces to $\sim$ 2, which
gives $\sim$ 3 in average.

The propagation speed of intensity blob is much higher comparing to
the local sound speed. Therefore, it can not be a simple slow
sausage pulse. One may compare the blob properties to new features
observed by Hinode; such as chromospheric jet-like structures
\citep{Katsukawa,Shibata,Nishizuka} and type II spicules \citep{De Pontieu1}. The observed jets
have different properties inside and outside sunspots. The
chromospheric jets observed in penumbral chromosphere have length of
1-4 Mm, width of 400 km and apparent rise velocity of $>$ 100 km
s$^{-1}$ \citep{Katsukawa}. The anemone jets observed outside
sunspots are 2-5 Mm length, 150-300 km width and have apparent
velocity of 10-20 km s$^{-1}$ \citep{Shibata}. The type II
spicules have life time of 10-150 s, apparent upward velocity of
50-150 km s$^{-1}$ and width of 200 km \citep{De Pontieu1}.
They are tallest reaching 5000 km or more in coronal holes, while in
quiet Sun regions they reach lengths of several Mm. The length and
apparent speed of our intensity blob do not coincide to neither of
these features; it is shorter than the observed jets, has different
upward speed and propagates as a pulse-like structure, not a jet.
Therefore, we argue that the blob represents either a fast sausage
pulse or slow sausage soliton.

Fast sausage pulse may propagate much faster than the local sound
speed for relatively larger external Alfv\'en speed \citep{edwin}. However, fast sausage waves are leaky for the long
wave-length limit. Suppose, that the radius of the tube where the
fast sausage pulse (or wave trains) propagates is $a$, then we get
$ka=2 \pi a/l=2\pi/6 \approx 1$, where $l$ is the characteristic
length of the pulse (here $l/a$ parameter is taken from observed
length to width ratio of the blob). The fast sausage waves are leaky
for this value of $ka$ (see Fig. 4 in \citet{edwin}).
Therefore, the pulse should vanish rapidly before it could propagate
upwards.

\begin{figure*}
\centering
\includegraphics[width=15cm]{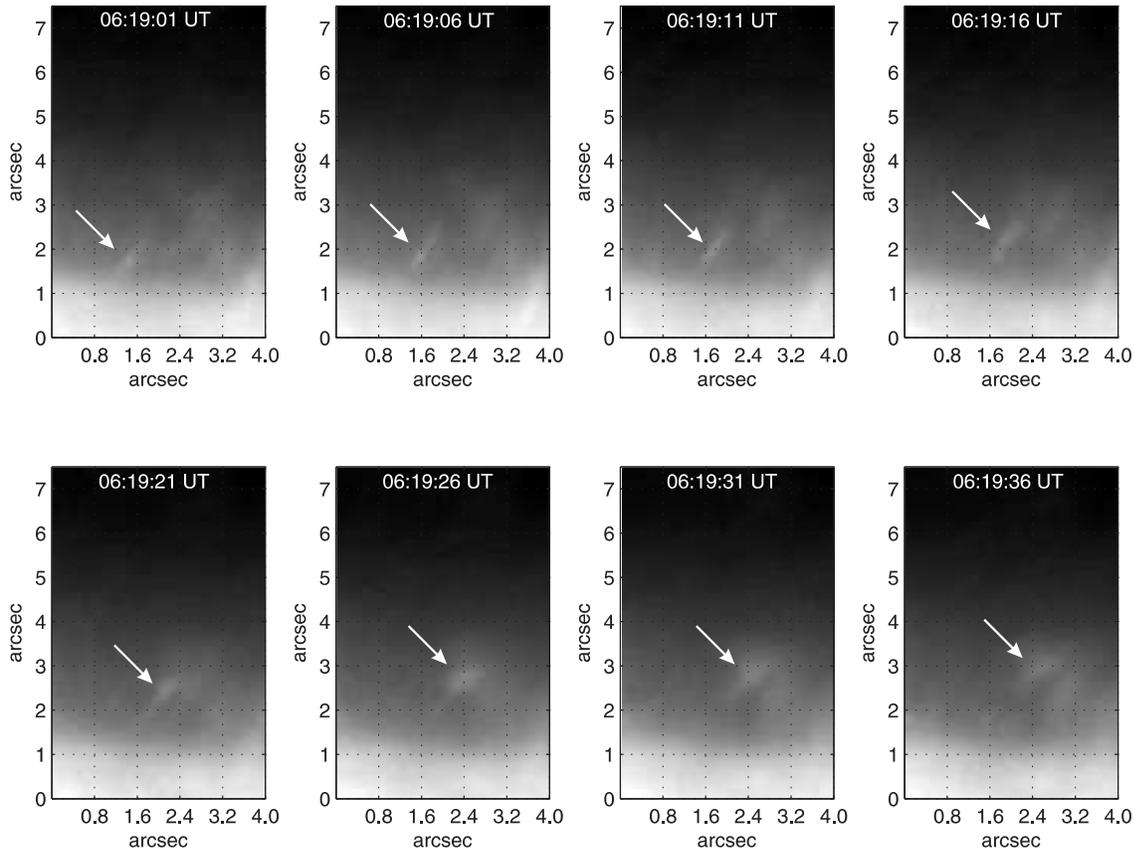}
\caption{8 consecutive images of time sequence in Ca II H line (left
to right and top to bottom). The intensity blob is located near
$(x,y)$=(1.4,1.8) point in the first image (upper left panel). The
blob propagates upward with mean apparent speed of 35 km s$^{-1}$.
We identify it with the slow sausage soliton propagating along
magnetic flux tube.  }
\label{FigGam}%
 \end{figure*}

On the other hand, the observed blob propagates without significant
change of relative amplitude and the form (at least in the first 5
images), which may rule out the possibility of fast sausage pulse.
The blob changes its shape in the last 3 images becoming wider,
however length to width ratio and relative amplitude remain more or
less similar. The blob width is $\sim$ 136 km at the lowest height
and increases up to $\sim$ 516 km at the highest height (Fig. 3,
lower panel). The observed broadening may reflect the magnetic tube
expansion with height (we will discuss it later).

Another scenario of the intensity blob propagation is a slow sausage
soliton, which is formed when non-linear steepening due to large
amplitude is balanced by wave dispersion. The soliton propagates
without significant changing of form and faster than the tube speed.
The soliton solution should satisfy the parameters, which can be
tested from observational properties of the pressure blob. The
soliton solution in structured magnetic field is well-studied
\citep{Roberts1,Roberts,Roberts2,Merzljakov,Merzljakov2,ruderman}. Therefore, there is no need to go for
detailed calculation of the parameters; we just use known
theoretical properties of slow sausage soliton and then
compare them with observations. A slow sausage soliton can be either surface or body solution depending on its structure inside the tube \citep{zhugzhda}. The quasi-homogeneous structure of observed blob suggests more surface than body soliton. Therefore, here we consider the slow surface sausage soliton, however the body solution can be also tested in the future.  

\section{Soliton solution}

Theoretical properties of sausage soliton are more easily obtained
for magnetic slabs rather than tubes. Numerical simulations of
solitary waves in magnetic tubes \citep{Ofman}
show the same properties of slow soliton as derived by analytical
calculations for a magnetic slab \citep{Roberts1}. Therefore, we consider a magnetic slab of width
$2x_0$ embedded in a magnetized environment. Let's suppose that the
magnetic field inside (outside) the slab is $B_0$ ($B_e$), the
density inside (outside) is $\rho_0$ ($\rho_e$) and the plasma
pressure inside (outside) is $p_0$ ($p_e$). The pressure balance
condition at the slab boundaries is $p_0+B^2_0/2\mu=p_e+B^2_e/2\mu$.
The characteristic wave speeds inside (outside) the slab are: the
Alfv\'en speed $V_A=B_0(\mu \rho_0)^{-1/2}$ ( $V_{Ae}=B_e(\mu
\rho_e)^{-1/2}$), the sound speed $c_s=(\gamma p_0/\rho_0)^{1/2}$
($c_{se}=(\gamma p_e/\rho_e)^{1/2}$) and the tube speed
$c^2_T={{c^2_s V^2_A}/( {c^2_s+ V^2_A})}$ ($c^2_{Te}={{c^2_{se}
V^2_{Ae}}/( {c^2_{se}+ V^2_{Ae}})}$). An important parameter of wave
propagation in magnetic slabs is
$m_e=\sqrt{{{(V^2_{Ae}-c^2_T)(c^2_{se}-c^2_T)}/
{[(V^2_{Ae}+c^2_{se})(c^2_{Te}-c^2_T)]}}}$, which plays the role of
perpendicular wave number outside the slab. The waves may propagate
in the slab only when $m^2_e >0$ (they are leaky if $m^2_e <0$).

The solution of slow sausage surface soliton in magnetic slabs can
be given by the following expression \citep{ruderman}:

\begin{equation}
\eta={{al^2}\over {l^2 + [z-s t]^2}},
\end{equation}
where $\eta$ is the displacement of the slab boundary, $a$ is the
maximal value of the displacement $\eta$ (i.e. the soliton
amplitude) and
\begin{equation}
s = c_T + {1\over 4}{{ab}\over {x_0}},\,\, l=4{{\kappa x_0}\over
{ab}}
\end{equation}
are the soliton speed and the spatial scale respectively.

The parameters $b$ and $\kappa$ are expressed as
\begin{equation}
b={{V^4_A[3c^2_s+(\gamma+1)V^2_A]}\over {2c_T(c^2_s + V^2_A)^2}},
\,\, \kappa={x_0\over 2}{\rho_{e0}\over \rho_0}{{c_Tc^2_s(c^2_T
-V^2_{Ae})}\over {m_eV^2_A(c^2_s + V^2_A)}}.
\end{equation}

Let us check if the observed parameters of intensity blob satisfy
the requirements of sausage soliton. The observed blob propagates in
the chromosphere, where the sound speed can be taken as $c_s$=10 km
s$^{-1}$. We assume the density ratio outside and inside the slab as
${\rho_{e}/\rho_{0}}$=0.9. Then, the propagation speed of the
soliton, $s$, is determined by the soliton relative amplitude,
$a/x_0$, and the Alfv\'en speed $V_A$ (see Eq. (2)). The observed
relative amplitude of the blob is estimated as $a/x_0$=0.2, then the
Alfv\'en speed stays as a free parameter. In order to obtain the
observed apparent propagation speed i.e. 35 km s$^{-1}$, the
Alfv\'en speed needs to be $\sim 70$ km s$^{-1}$.

Another important parameter of the sausage soliton is its length to
width ratio. Observations show that the blob has elongated form; its
mean length is approximately 3 times larger than width. Then the
parameter associated to the soliton length is $l\approx 6 x_0$. This
will be achieved when $m_e \ll 1$, which in turn requires
$c_{se}\rightarrow c_{T}$ or $c_{se}\rightarrow c_{s}$ (as the
Alfv\'en speed is much higher than the sound speed). Thus, the
soliton may have the observed elongated shape if electron
temperature inside and outside the tube is approximately similar.

\section{Discussion}

Brief conclusion of the previous section is that the observed
intensity blob may represent a slow sausage soliton in chromospheric
magnetic tube, which has the Alfv\'en speed of $\sim 70$ km s$^{-1}$
and the temperature balance with surroundings. Note, that the both
requirements are quite typical to the chromosphere. Possible
inclination of the tube along the line of sight may cause additional
correction to the estimated Alfv\'en speed. $35^0$ inclination leads
to the blob propagation speed of $\sim 42$ km s$^{-1}$, which then
cause the slight increase of required Alfv\'en speed.

\begin{figure}
   \includegraphics[width=10cm]{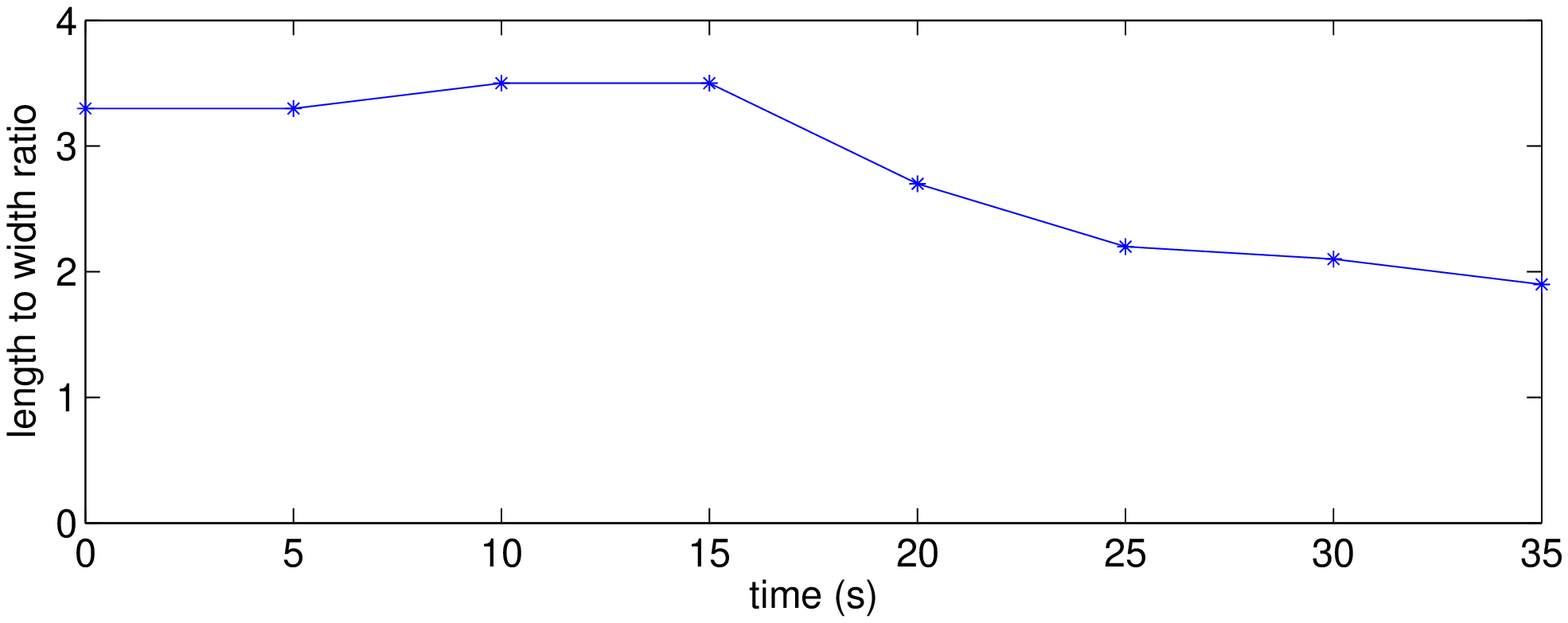}
   \includegraphics[width=10cm]{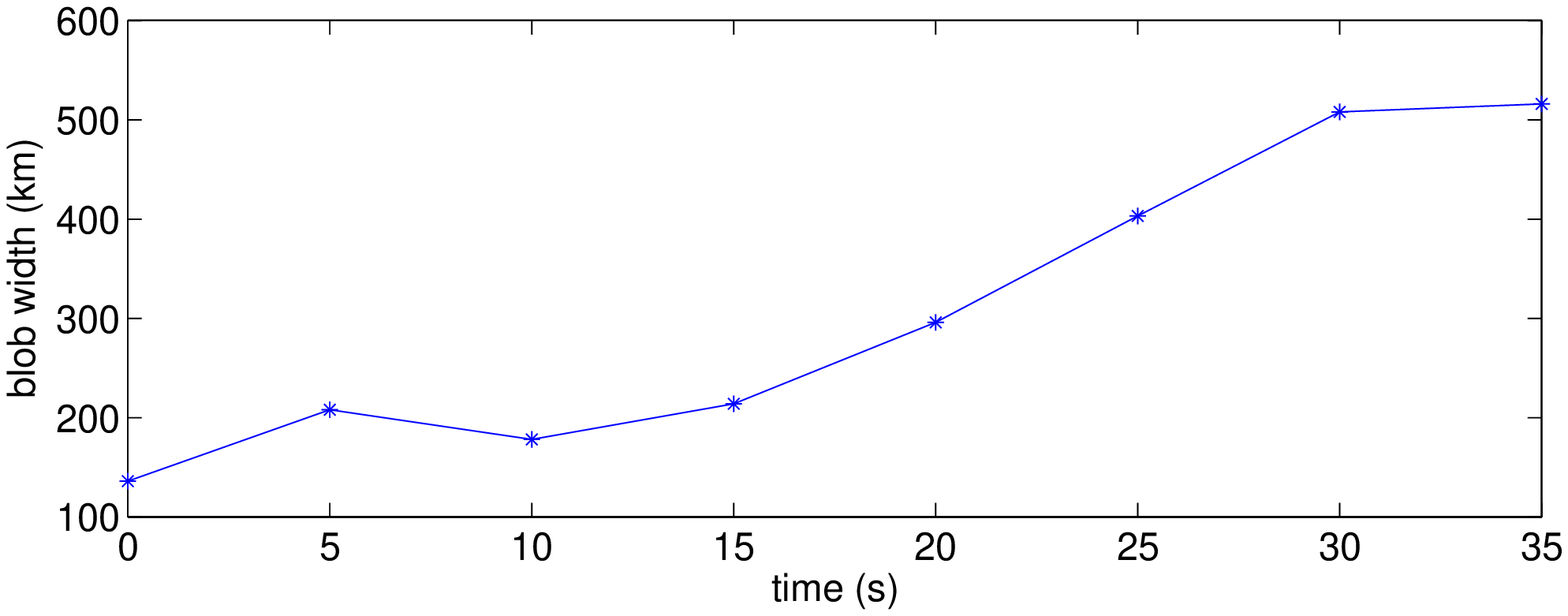}
      \caption{Length to width ratio (upper plot) and width (lower plot) of intensity blob vs time from the sequence of Fig. 2.
              }
         \label{FigVibStab}
   \end{figure}

It is interesting to note that the blob form remains almost
unchanged in the first 4 images of Fig. 2. However, in the last four
images the width of the blob is significantly increased. Fig. 3
(lower panel) shows the variation of the blob width with time. The
gradual increase of the blob width probably is due to the expansion
of the magnetic tube with height due to the stratification of the
solar atmosphere. In thin flux tube approximation, magnetic field
strength varies as $B_0(z)=B_0(0)\exp{(-z/2h)}$ in the simplest case
of isothermal atmosphere, where $h$ is the scale height.
Conservation of magnetic flux yields $B_0(z)A(z)=const$, where
$A(z)$ is the tube cross section. Then, the dependence of the tube
diameter on height should be as $\sim \exp{(z/4h)}$. This dependence
can be used up to 1200 km, where the thin flux tube approximation is
valid \citep{Hasan}. The scale height of $\sim$ 220 km
(estimated for the sound speed of 10 km s$^{-1}$) yields an increase
of the tube diameter by $\sim 2.2$ times between 500 and 1200 km
heights. The observed width at 500 km and 1200 km is 136 and 300 km
respectively. The ratio between the two parameters  gives exactly
suggested value. Thus, the blob propagates along the magnetic tube,
which is expended upwards as modeled by the stratified atmosphere.
The rapid broadening of the blob in last images (i.e. higher
heights) may correspond to the rapid increase of the tube
cross-section (see Fig. 1 in Hasan et al. 2003).

The blob begins to disappear after the height of $\sim$ 1700 km
probably due to the changed conditions for soliton formation. Due to
the mathematical difficulties, all known theoretical properties of
slow surface sausage soliton were calculated for the magnetic tubes
with constant cross section. Therefore, it is unclear what happens
when the soliton propagates along the tubes with varying cross
section. Intuitively, one may suppose that the soliton parameters
also slowly vary during the propagation. It also should be mentioned
that the soliton solution, which we use to model the blob
propagation, was obtained without taking into account the
stratification, which is important in this part of solar atmosphere.
These problems need further detailed study in theoretically and
numerically.

The length and apparent speed of intensity blob is quite different
from chromospheric jet-like structures \citep{Katsukawa,Shibata,Nishizuka} and type II spicules \citep{De Pontieu1}. Interpretation of the blob as a plasmoid
propagating after a magnetic reconnection can be also ruled out as
no explosive event is detected in upper photosphere during the
observations. If magnetic reconnection took place in
sub-photospheric layers, then the plasmoid should have much higher
density than it is observed. The propagation speed of the blob can
be modeled by transverse kink or fast sausage pulses as well.
The second possibility is unlikely to occur as the observed spatial scale leads to the leaky regime of
fast sausage waves. The first possibility needs further discussion as a kink pulse may lead to the intensity enhancement
in inclined magnetic tubes \citep{cooper}. However, it requires very large amplitude and the pulse may have significantly curved form, which is not observed. Therefore, the kink pulse is unlikely to be the reason of the intensity enhancement.

Following to the discussion above, it seems that the slow sausage
scenario has strong background. We suggest that this is the first
observation of sausage soliton in the solar atmosphere. We believe
that careful analysis of SOT time series will reveal other similar
cases, which may enhance the interest to the soliton physics in the
solar atmosphere.

Of additional importance for the quantitative interpretation of the
observed phenomenon as a propagating solitary wave would be taking
into account of the plasma partial ionization effects in the solar
chromosphere. The presence of even a small amount of neutral atoms
in plasma is known to change significantly its dynamical and
physical properties \citep{Braginskii,Khodachenko1,Khodachenko2}. Different interaction of electrons, ions
and neutral atoms with the magnetic field and each other causes the
main specifics of the partially ionized plasma MHD, which differs
significantly from the fully ionized plasma case. The inclusion of
the ion-neutral collision effects into the scope of the proposed
interpretation requires a special theoretical study of solitary
waves behaviour in partially ionized plasmas which represents a
subject for future work.

It would be interesting to search analogy of soliton-like formations
on the solar disc. Possibly, chromospheric bright grains \citep{lites}, which were often associated to the shocks, represent
soliton formations in magnetic tubes. Future detailed study is
necessary to identify these features.

\section{Conclusions}

\begin{enumerate}
\item Time series of Ca II H line obtained by Hinode/SOT at the solar limb shows upward propagating intensity blob.
The blob appears at 500-600 km height above the surface and reaches
to the height of $\sim$ 1700 km after 35 s. Therefore, the mean
apparent propagation speed is 35 km s$^{-1}$. The blob has elongated
form and the length to width ratio is $\sim$ 3 in average. The
length to width ratio, the relative intensity and the propagation
speed change slightly during the propagation.
\item The observed parameters fit with theoretically expected properties of slow sausage soliton propagating along a
magnetic flux tube, which has the Alfv\'en speed of $\sim$ 70 km
s$^{-1}$ and is in temperature balance with surroundings (note, that
an inclination of the tube along the line of sight may slightly
increase the value of Alfv\'en speed). Therefore, we suggest that
this is the first observational evidence of slow sausage soliton in
the solar atmosphere.
\item The width of the blob increases with height, which coincides with the expected expansion of magnetic tubes in
the stratified atmosphere.
\end{enumerate}

\section*{Acknowledgments}

This work was supported by the Austrian Fond zur F\"orderung der
wissenschaftlichen Forschung (project P21197-N16) and the Georgian National Science
Foundation grant GNSF/ST09/4-310. Hinode is a
Japanese mission developed and launched by ISAS/JAXA, with NAOJ as
domestic partner and NASA and STFC (UK) as international partners.
It is operated by these agencies in co-operation with ESA and NSC
(Norway).

\bsp

\label{lastpage}


\begin{thebibliography}{99}

\bibitem[\protect\citeauthoryear{Baird}{1981}]{b1} Baird S.R., 1981,
ApJ, 245, 208


\bibitem[\protect\citeauthoryear{Ballai et al.}{2003}]{ballai}Ballai, I., Thelen, J.C. \& Roberts, B. 2003,
      A\&A, 404, 701

\bibitem[\protect\citeauthoryear{Braginskii}{1965}]{Braginskii}Braginskii, S.I., 1965, Transport processes in plasma, In: Reviews of plasma physics, V.1 (Consultants Bureau, New York)

\bibitem[\protect\citeauthoryear{Cooper et al.}{2003}]{cooper}Cooper, F.C., Nakariakov, V.M. \& Tsiklauri, D., 2003, A\&A, 397, 765


\bibitem[\protect\citeauthoryear{De Pontieu et al.}{2007a}]{De Pontieu}De Pontieu, B. et al., 2007a, Science, 318, 1574

\bibitem[\protect\citeauthoryear{De Pontieu et al.}{2007b}]{De Pontieu1}De Pontieu, B. et al., 2007b, PASJ,
59, S655

\bibitem[\protect\citeauthoryear{Edwin \& Roberts}{1983}]{edwin}Edwin, P.M. \& Roberts, B. 1983, Solar Phys., 88, 179

\bibitem[\protect\citeauthoryear{Erd\'elyi \& Fedun}{2006}]{erdelyi}Erd\'elyi, R. \& Fedun, V. 2006,
      Phys. of Plasmas, 13, 032902

\bibitem[\protect\citeauthoryear{Harrison}{1997}]{Harrison}Harrison, R.A., 1997, Solar Phys., 175, 467

\bibitem[\protect\citeauthoryear{Hasan et al.}{2003}]{Hasan}Hasan, S.S., Kalkofen, W., van
Ballegooijen, A.A. and Ulmschneider, P. 2003, ApJ, 585, 1138

\bibitem[\protect\citeauthoryear{Katsukawa et al.}{2007}]{Katsukawa}Katsukawa, Y. et al. 2007, Science, 318, 1594

\bibitem[\protect\citeauthoryear{Khodachenko \& Zaitsev}{2002}]{Khodachenko1}Khodachenko M.L. \& Zaitsev V.V., 2002, Astrophys. \& Space Science, 279, 389

\bibitem[\protect\citeauthoryear{Khodachenko et al.}{2004}]{Khodachenko2}Khodachenko M.L., Arber, T.D., Rucker, H.O. \& Hanslmeier, A., 2004, A\&A, 422, 1073

\bibitem[\protect\citeauthoryear{Kukhianidze et al.}{2006}]{kukhianidze}Kukhianidze, V., Zaqarashvili, T.V. \& Khutsishvili, E., 2006,
      A\&A, 449, L35

\bibitem[\protect\citeauthoryear{Lites et al.}{1999}]{lites}Lites, B. W., Rutten, R. J. \& Berger, T. E., 1999, ApJ, 517, 1013

\bibitem[\protect\citeauthoryear{Jess et al.}{2009}]{Jess}Jess, D.B. et al. 2009, Science, 323, 1582

\bibitem[\protect\citeauthoryear{Merzljakov \& Ruderman}{1985}]{Merzljakov}Merzljakov, E. G. \& Ruderman, M.S. 1985,
      Solar Phys., 95, 51

\bibitem[\protect\citeauthoryear{Merzljakov \& Ruderman}{1986}]{Merzljakov2}Merzljakov, E. G. \& Ruderman, M.S. 1986,
      Solar Phys., 103, 259

\bibitem[\protect\citeauthoryear{Nakariakov \& Roberts}{1999}]{Nakariakov}Nakariakov, V.M. \& Roberts, B. 1999,
      Phys. Let. A, 254, 314

\bibitem[\protect\citeauthoryear{Nishizuka et al.}{2008}]{Nishizuka}Nishizuka, N., Shimizu, M., Nakamura,
T., Otsuji, K., Okamoto, T. J., Katsukawa, Y. and Shibata, K., 2008,
ApJ, 683, L83

\bibitem[\protect\citeauthoryear{Ofman \& Davila}{1997}]{Ofman}Ofman, L. \& Davila, J.M. 1997, ApJ,  476, 357

\bibitem[\protect\citeauthoryear{Porter \& Dere}{1991}]{Porter}Porter, J., \& Dere, K. 1991, ApJ, 370, 775

\bibitem[\protect\citeauthoryear{Roberts \& Mangeney}{1982}]{Roberts1}Roberts, B. \& Mangeney, A. 1982,
      MNRAS, 198, 7

\bibitem[\protect\citeauthoryear{Roberts}{1985}]{Roberts}Roberts, B. 1985, Phys. Fluids, 28, 3280

\bibitem[\protect\citeauthoryear{Roberts}{1987}]{Roberts2}Roberts, B. 1987, ApJ, 318, 590

\bibitem[\protect\citeauthoryear{Ruderman}{2003}]{ruderman}Ruderman, M.S. 2003, Turbulence, Waves, and Instabilities in the Solar
Plasma ed. Erd\'elyi et al. (Kluwer Academic Publisher, Dordrecht)
239

\bibitem[\protect\citeauthoryear{Ryutova \& Hagenaar}{2007}]{ryutova}Ryutova, M. \& Hagenaar, H. 2007, Solar Phys., 246, 281

\bibitem[\protect\citeauthoryear{Sahyouni et al.}{1988}]{Sahyouni}Sahyouni, W., Zheliazkov, I. \& Nenovski, P. 1988, Solar Phys., 115, 17

\bibitem[\protect\citeauthoryear{Shibata et al.}{2007}]{Shibata}Shibata, K. et al. 2007, Science, 318,
1591

\bibitem[\protect\citeauthoryear{Tsuneta et al.}{2008}]{Tsuneta}Tsuneta, S. et al. 2008, Solar Phys., 259, 167

\bibitem[\protect\citeauthoryear{Zaqarashvili et al.}{2007}]{zaqarashvili}Zaqarashvili, T.V., Khutsishvili, E., Kukhianidze, V. \& Ramishvili, G. 2007,
      A\&A, 474, 627

\bibitem[\protect\citeauthoryear{Zaqarashvili \& Erd\'elyi}{2009}]{zaqarashvili1}Zaqarashvili, T.V. \& Erd\'elyi, R. 2009,
      Space Science Rev., 149, 355
      
\bibitem[\protect\citeauthoryear{Zhugzhda \& Nakariakov}{1997}]{zhugzhda}Zhugzhda, Y.D. \& Nakariakov, V.M., 1997, Phys. Let. A, 233, 413


\end{thebibliography}
\end{document}